\documentclass[seceq,supplement]{ptptex}

\usepackage{graphicx}
\usepackage{wrapft}

    \usepackage{bm}

\newcommand\beq{\begin{equation}}
\newcommand\eeq{\end{equation}}
\newcommand\beqa{\begin{eqnarray}}
\newcommand\eeqa{\end{eqnarray}}
\newcommand{\nn}{\nonumber\\}

\newcommand{\ma}{m_i}

\newcommand{\mb}{m_j}

\newcommand{\cca}{\mathbf{v}_i}

\newcommand{\ca}{{v}_i}

\newcommand{\ccb}{\mathbf{v}_j}

\newcommand{\cb}{{v}_j}

\newcommand{\JJ}{\mathbf{Q}_{ij}}
\newcommand{\JJw}{{\mathbf{Q}}_{ij}^-}

\newcommand{\J}{{Q}_{ij}}

\newcommand{\kk}{\widehat{\bm{\sigma}}}

\newcommand{\wwa}{\bm{\omega}_i}

\newcommand{\wwb}{\bm{\omega}_j}

\newcommand{\wa}{{\omega}_i}

\newcommand{\wb}{{\omega}_j}

\newcommand{\Ia}{I_i}

\newcommand{\Ib}{I_j}

\newcommand{\da}{\sigma_i}

\newcommand{\db}{\sigma_j}

\newcommand{\ds}{\sigma}

\newcommand{\dab}{\sigma_{ij}}

\newcommand{\x}{\times}

\newcommand{\gh}{\mathbf{v}_{ij}}

\newcommand{\g}{{v}_{ij}}

\newcommand{\een}{\alpha_{ij}}

\newcommand{\esn}{\alpha}

\newcommand{\eet}{\beta_{ij}}

\newcommand{\est}{\beta}

\newcommand{\en}{\widetilde{\alpha}_{ij}}

\newcommand{\et}{\widetilde{\beta}_{ij}}

\newcommand{\mab}{m_{ij}}

\newcommand{\qab}{\kappa_{ij}}

\newcommand{\qa}{\kappa_{i}}

\newcommand{\qb}{\kappa_{j}}

\newcommand{\q}{\kappa}

\newcommand{\fa}{f_{i}}

\newcommand{\fab}{f_{ij}^{(2)}}

\newcommand{\ffab}{\bar{f}_{ij}^{(2)}}

\newcommand{\fat}{f_{i}^\text{tr}}

\newcommand{\far}{f_{i}^\text{rot}}

\newcommand{\fbr}{f_{j}^\text{rot}}

\newcommand{\Tat}{T_{i}^\text{tr}}

\newcommand{\Tbt}{T_{j}^\text{tr}}

\newcommand{\Tt}{T^\text{tr}}

\newcommand{\Tar}{T_{i}^\text{rot}}

\newcommand{\Tbr}{T_{j}^\text{rot}}

\newcommand{\Tr}{T^\text{rot}}

\newcommand{\Qab}{J_{ij}}

\newcommand{\Iab}{\mathcal{J}_{ij}}

\newcommand{\na}{n_i}

\newcommand{\nb}{n_j}

\newcommand{\zabt}{\xi_{ij}^\text{tr}}

\newcommand{\zt}{\xi^\text{tr}}

\newcommand{\zabr}{\xi_{ij}^\text{rot}}

\newcommand{\zr}{\xi^\text{rot}}

\newcommand{\al}{i}

\newcommand{\be}{j}

\newcommand{\tr}{\text{tr}}

\newcommand{\rot}{\text{rot}}

\newcommand{\wwwa}{\mathbf{w}_i}
\newcommand{\wwwb}{\mathbf{w}_j}
\newcommand{\wwwab}{\mathbf{w}_{ij}}
\newcommand{\SSab}{\mathbf{S}_{ij}}
\newcommand{\Sab}{{S}_{ij}}
\newcommand{\chiab}{{\chi}_{ij}}

\title{
Energy production rates in fluid mixtures of inelastic rough hard spheres%
}

\author{
Andr\'es \textsc{Santos},$^{1,}$\footnote{E-mail: andres@unex.es} Gilberto M.  \textsc{Kremer}$^{2,}$\footnote{E-mail: kremer@fisica.ufpr.br}
and Vicente \textsc{Garz\'o}$^{1,}$\footnote{E-mail: vicenteg@unex.es}%
}

\inst{
$^1$Departamento de F\'{\i}sica, Universidad de
Extremadura, E-06071 Badajoz, Spain\\
$^2$Departamento de F\'{\i}sica, Universidade Federal do Paran\'a, Curitiba, Brazil
}

\abst{
 The aim of this work is to explore the combined effect of polydispersity and roughness on the partial energy production rates  and on the total cooling rate of a granular fluid mixture. We  consider a  mixture of inelastic rough hard spheres of different number densities, masses, diameters, moments of inertia, and mutual coefficients of normal  and tangential restitution. Starting from the first equation of the BBGKY hierarchy, the collisional energy production rates  associated with the translational and rotational temperatures ($T_i^\text{tr}$ and $T_i^\text{rot}$) are expressed in terms of two-body average values. Next, those average values are estimated by assuming a velocity distribution function based on maximum-entropy
arguments, allowing us to express the energy production rates and the total cooling rate  in terms of the partial temperatures and the parameters of the mixture. Finally, the results are applied to the homogeneous cooling state  of a binary mixture and the influence of inelasticity and roughness on the temperature ratios $T_1^\text{tr}/T_1^\text{rot}$, $T_2^\text{tr}/T_1^\text{tr}$, and $T_2^\text{rot}/T_1^\text{rot}$ is analyzed.
}

\begin{document}

\maketitle

\section{Introduction}

A granular fluid is usually modeled as a system of identical, inelastic smooth hard spheres with a constant coefficient of normal restitution $\alpha$.  Despite its simplicity, this model has been useful to capture the basic properties of granular flows.\cite{G03} On the other hand, the model can be made closer to reality by introducing more ingredients, such as  coefficients of normal restitution depending on the impact velocity,\cite{BSSP04} presence of an interstitial fluid,\cite{XVKL09} non-spherical shapes,\cite{CZMP09} polydispersity,\cite{G08} or roughness.\cite{MSS04} Of course, the few citations in the preceding sentence are just  representative of many works reporting features not accounted for by the simple monodisperse smooth-sphere model.

In this paper we will focus on the two latter ingredients, namely polydispersity  and roughness. These properties are especially relevant, not only because beads and grains are unavoidably polydisperse and rough, but also because any of these ingredients unveils an inherent breakdown of energy equipartition in granular fluids, even in homogeneous and isotropic states. In the case of multi-component granular fluids, most of the studies  have considered the inelastic smooth-sphere description. Some of the problems addressed include non-equipartition in homogeneous states,\cite{GD99,MG02,BT02,DHGD02} Navier--Stokes transport coefficients,\cite{JM89,GD02,SGNT06,GDH07,GHD07} and segregation phenomena.\cite{HQL01,JY02,K04,BRM05,G08a}

Concerning the case of inelastic rough spheres, most of the works we are aware of are restricted to monodisperse systems.\cite{JR85,LS87,C89,L91,LN94,GS95,L96,ZVPSH98,HZ98,ML98,LHMZ98,HHZ00,AMZ01,MHN02,CLH02,JZ02,PZMZ02,L95,GNB05,Z06,BPKZ07}
Analogously to what happens with the coefficient of normal restitution $\alpha$, the simplest model accounting for friction during collisions assumes a \emph{constant} coefficient of tangential restitution $\beta$. While $\alpha$ is a positive quantity smaller than or equal to 1 (the value $\alpha=1$ corresponding to elastic spheres), the parameter $\beta$ lies in the range between $-1$ (perfectly smooth spheres) to $1$  (perfectly rough spheres). Except for $\alpha=1$ and $ \beta=\pm 1$, the total kinetic energy is not conserved in a collision. Some of the early attempts to develop a kinetic theory for rough spheres were carried out by Jenkins and Richman\cite{JR85} and Lun and Savage,\cite{LS87} who applied their approaches to the simple shear flow problem. The influence of roughness in shear flows has also been studied by several authors,\cite{C89,L91,LN94,L96,ZVPSH98,JZ02,PZMZ02} usually assuming that the spheres are nearly smooth and nearly elastic. In an extensive paper,\cite{GS95} Goldshtein and Shapiro obtained the collisional energy production rates associated with the translational and rotational degrees of freedom by using Maxwellian forms for the distribution functions. The result was applied to the evaluation of the ratio between the translational ($\Tt$) and rotational ($\Tr$) temperatures in the homogeneous cooling state (HCS). The time evolution of the ratio $\Tt/\Tr$ towards its HCS asymptotic value has been widely analyzed, both theoretically and by means of molecular dynamics, by Luding, Zippelius, and co-workers.\cite{HZ98,ML98,LHMZ98,HHZ00,AMZ01,CLH02,Z06} Other studies involving roughness include vibration with rough walls,\cite{L95} a micropolar  fluid model for granular flows on a slope,\cite{MHN02}  derivation of hydrodynamic constitutive equations from the Boltzmann equation for nearly smooth, nearly elastic granular gases,\cite{GNB05} and correlations between the rotation axis and the translational direction.\cite{BPKZ07}

The studies about multi-component rough-sphere systems are much scarcer. To the best of our knowledge, only the case of a fixed particle immersed in a bath of thermalized point particles has been addressed.\cite{VT04,PTV07,CP08} On the other hand, the general case of a polydisperse system made of mobile particles of different coefficients of normal and tangential restitution ($\een$ and $\eet$) has not been studied yet. In this paper, we address one of the basic aspects of the problem, namely those related to the partition of the total energy. In order to characterize the effect of collisions on energy partition, we focus on the partial productions rates $\zabt$ and $\zabr$ measuring the rate of change of the translational and rotational kinetic energies, respectively, of particles of component $\al$ due to collisions with particles of component $\be$. A combination of $\zabt$ and $\zabr$ gives the total cooling rate $\zeta$ of the mixture.

Starting from the collisional rules worked out in Section \ref{sec2}, the collisional rates of change of  momentum (linear and angular) and energy (translational and rotational) are expressed as linear combinations of two-body average values in Section \ref{sec3}. These averages are evaluated in terms of the partial temperatures $\Tat$, $\Tar$, $\Tbt$, and $\Tbr$ by assuming a maximum-entropy two-body distribution in Section \ref{sec4}. The expressions for $\zabt$, $\zabr$, and $\zeta$  obtained in Section \ref{sec5} extend previous results derived  for monodisperse rough spheres\cite{GS95,LHMZ98} and for polydisperse smooth spheres.\cite{GD99} An application to the HCS of a binary mixture is carried out in Section \ref{sec6} with some illustrative examples. Finally, some concluding remarks are presented in Section \ref{sec7}.

\section{Collisional rules}
\label{sec2}

\begin{figure}[tbp]
\centerline{\includegraphics[width=.5\columnwidth]{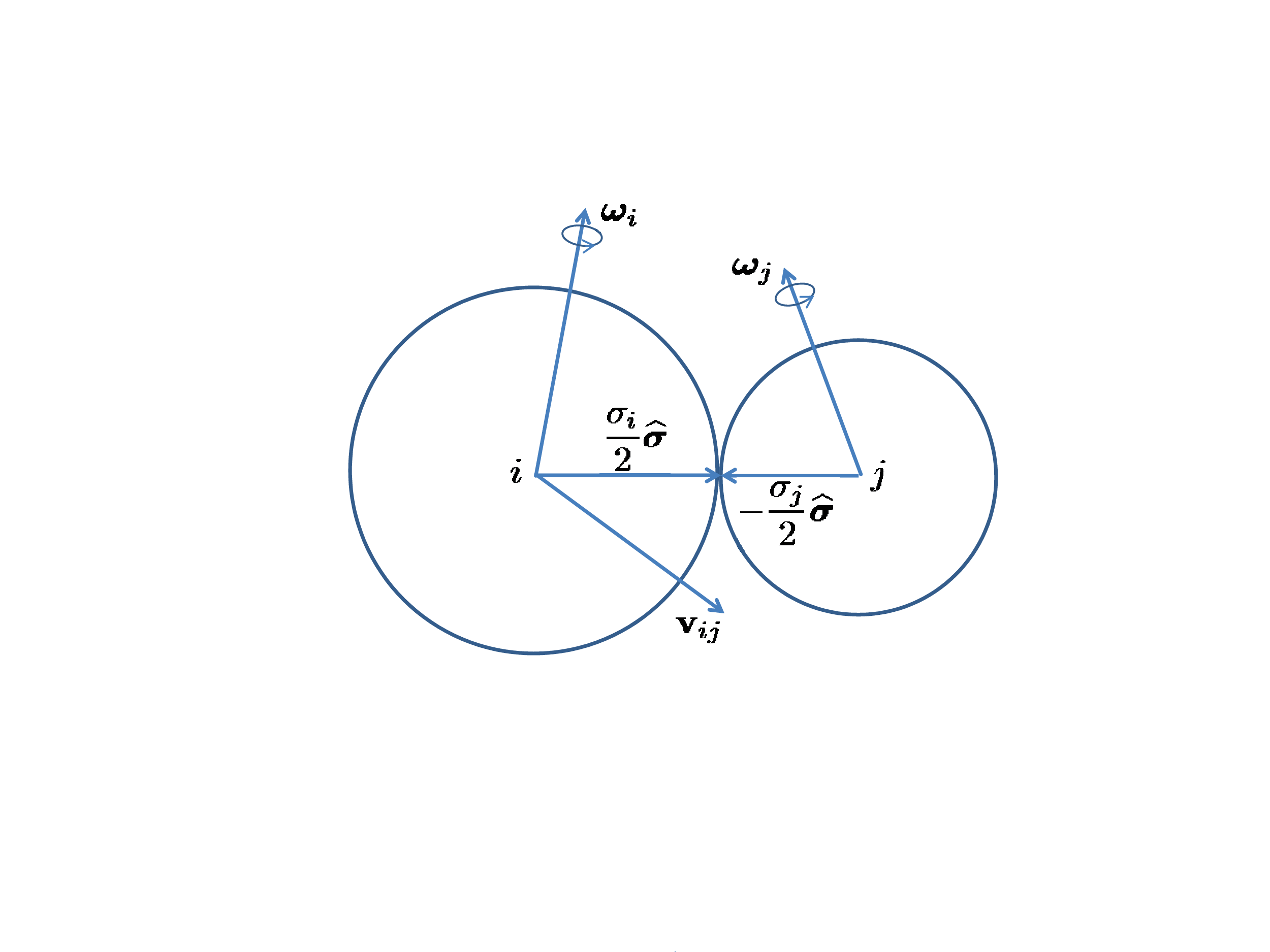}}
\caption{Sketch of the pre-collisional quantities of spheres $i$ and $j$ in the frame of reference solidary with particle $j$.}
\label{sketch}
\end{figure}

Let us consider the collision between two hard spheres of masses $\ma$ and $\mb$, diameters $\da$ and $\db$, and moments of inertia $\Ia$ and $\Ib$. The latter two quantities can be equivalently characterized by the dimensionless parameters
\beq
\qa\equiv\frac{4\Ia}{\ma\da^2},\quad \qb\equiv\frac{4\Ib}{\mb\db^2}.
\label{kappa}
\eeq
The value of $\qa$ depends on the mass distribution within the sphere and runs from the extreme values $\qa=0$ (mass concentrated on the center) to $\qa=\frac{2}{3}$ (mass concentrated on the surface). If the mass is uniformly distributed, then $\qa=\frac{2}{5}$.
Let us denote by $\gh=\cca-\ccb$ the pre-collisional relative velocity of the center of mass of both spheres and by $\wwa$ and $\wwb$ the respective pre-collisional angular velocities. This is sketched in Fig.\ \ref{sketch}, where $\kk\equiv (\mathbf{r}_j-\mathbf{r}_i)/|\mathbf{r}_j-\mathbf{r}_i|$ is the unit vector pointing from the center of $i$ to the center of $j$. The velocities of  the points of the spheres which
are in contact during the collision are
\beq
\wwwa=\cca-\frac{\da}{2}\kk\x\wwa,\quad \wwwb=\ccb+\frac{\db}{2}\kk\x\wwb,
\label{w}
\eeq
the corresponding relative velocity being
\beq
\wwwab=\gh-\kk\x\SSab,\quad \SSab\equiv \frac{\da}{2}\wwa+ \frac{\db}{2}\wwb.
\label{Sab}
\eeq

Post-collisional velocities will be denoted by primes. Conservation of linear and angular momenta yields\cite{Z06}
\beq
\ma\cca'+\mb\ccb'=\ma\cca+\mb\ccb,
\label{momentum}
\eeq
\beq
\Ia\wwa'-\ma\frac{\da}{2}\kk\x\cca'=\Ia\wwa-\ma\frac{\da}{2}\kk\x\cca,\quad \Ib\wwb'+\mb\frac{\db}{2}\kk\x\ccb'=\Ib\wwb+\mb\frac{\db}{2}\kk\x\ccb.
\label{angular}
\eeq
Equations \eqref{momentum} and \eqref{angular} imply that
\beq
\cca'=\cca-\frac{1}{\ma}\JJ,\quad \ccb'=\ccb+\frac{1}{\mb}\JJ,
\label{13b}
\eeq
\beq
\wwa'=\wwa-\frac{\da}{2\Ia}\kk\x\JJ, \quad \wwb'=\wwb-\frac{\db}{2\Ib}\kk\x\JJ,
\label{14}
\eeq
where $\JJ$ is the impulse exerted by particle $i$ on particle $j$.
Therefore,
\beq
\gh'=\gh-\frac{1}{\mab}\JJ,\quad \wwwab'=\wwwab-\frac{1}{\mab}\JJ+\frac{1}{\mab\qab}\kk\x\left(\kk\x\JJ\right),
\label{wwwab'}
\eeq
where
\beq
\mab\equiv \frac{\ma\mb}{\ma+\mb},\quad \qab\equiv \qa\qb\frac{\ma+\mb}{\qa\ma+\qb\mb}
\label{21}
\eeq
are the reduced mass and a sort of reduced inertia-moment parameter, respectively.

To close the collisional rules, we need to express $\JJ$ in terms of the pre-collisional velocities and the unit vector $\kk$. To that end, we relate the normal (i.e., parallel to $\kk$) and tangential (i.e., orthogonal to $\kk$) components of the relative velocities $\wwwab$ and $\wwwab'$ by
\beq
\kk\cdot \wwwab'=-\een \kk\cdot \wwwab,\quad \kk\x \wwwab'=-\eet \kk\x \wwwab.
\label{restitution}
\eeq
Here, $\een$ and $\eet$ are the  coefficients of normal and tangential restitution, respectively. The former coefficient ranges from $\een=0$ (perfectly inelastic particles) to $\een=1$ (perfectly elastic particles), while the latter runs from $\eet=-1$ (perfectly smooth particles) to $\eet=1$ (perfectly rough particles).
Inserting the second equality of Eq.\ \eqref{wwwab'} into Eq.\ \eqref{restitution} one simply gets $\kk\cdot\JJ=\en \kk\cdot\wwwab$ and $\kk\x\JJ=\et \kk\x\wwwab$, where
\beq
\en\equiv\mab\left(1+\een\right),\quad\et\equiv\frac{\mab\qab}{1+\qab}\left(1+\eet\right).
\label{20}
\eeq
Therefore,
\beq
\JJ=\en (\gh\cdot\kk)\kk+\et\left[\gh-\kk\x\SSab-(\gh\cdot\kk)\kk\right],
\label{15}
\eeq
where use has been made of the mathematical property $\kk\x(\kk \x\mathbf{A})=(\kk\cdot \mathbf{A})\kk-\mathbf{A}$.  Note that in the special case of perfectly smooth spheres ($\eet=-1$) one has $\et=0$, so that $\kk\x\JJ=\mathbf{0}$. In that case, according to Eq.\ \eqref{14}, the angular velocities of the two colliding spheres are unaltered by the collision.

The total kinetic energy before collision is
\beq
E_{ij}=\frac{\ma}{2}\ca^2+\frac{\mb}{2}\cb^2+\frac{\Ia}{2}\wa^2+\frac{\Ib}{2}\wb^2.
\label{Z2}
\eeq
It can be checked (see the Appendix) that
\beqa
E_{ij}'-E_{ij}&=&
-\frac{\mab}{2}\frac{\qab}{1+\qab}\left(1-\eet^2\right)\left[(\kk\x\gh)^2+(\kk\x\SSab)^2+
2(\kk\x\gh)\cdot\SSab\right]\nn
&&-\frac{\mab}{2}\left(1-\een^2\right)(\kk\cdot\gh)^2.
\label{29}
\eeqa
The right-hand side is a negative definite quantity. Thus, we observe that energy is conserved only if the particles are elastic  ($\een=1$)  and  either perfectly smooth ($\eet=-1$) or perfectly rough ($\eet=1$). Otherwise, $E_{ij}'<E_{ij}$ and kinetic energy is dissipated upon collisions.

Equations \eqref{13b}, \eqref{14}, and \eqref{15} give the \emph{direct} collisional rules. The \emph{restituting} collisional rules are
\beq
\cca''=\cca-\frac{1}{\ma}\JJw,\quad \ccb''=\ccb+\frac{1}{\mb}\JJw,
\label{13brest}
\eeq
\beq
\wwa''=\wwa-\frac{\da}{2\Ia}\kk\x\JJw, \quad \wwb''=\wwb-\frac{\db}{2\Ib}\kk\x\JJw,
\label{14rest}
\eeq
where
\beq
\JJw=\frac{\en}{\een} (\gh\cdot\kk)\kk+\frac{\et}{\eet}\left[\gh-\kk\x\SSab-(\gh\cdot\kk)\kk\right].
\label{15rest}
\eeq
Here the double primes denote pre-collisional quantities giving rise to unprimed quantities as post-collisional values.
 The modulus of the Jacobian of the transformation between pre- and post-collisional velocities is
\beq
\left|\frac{\partial(\cca',\wwa',\ccb',\wwb')}{\partial(\cca,\wwa,\ccb,\wwb)}\right|=
\left|\frac{\partial(\cca,\wwa,\ccb,\wwb)}{\partial(\cca'',\wwa'',\ccb'',\wwb'')}\right|={\een\eet^2}.
\label{Jacob}
\eeq

\section{Collisional rates of change}
\label{sec3}
Let $\fab(\mathbf{r}_i,\cca,\wwa;\mathbf{r}_j,\ccb,\wwb;t)$ be the two-body distribution function with the normalization condition
\beq
\int d\mathbf{r}_\al\int d\cca\int d\wwa\int d\mathbf{r}_\be\int d\ccb\int d\wwb \fab(\mathbf{r}_i,\cca,\wwa;\mathbf{r}_j,\ccb,\wwb;t)=N_\al N_\be,
\label{III.0}
\eeq
$N_\al$ being the number of spheres of component $\al$.
The  one-body distribution function is
\beq
\fa(\mathbf{r}_i,\cca,\wwa;t)=N_\be^{-1}\int d\mathbf{r}_j\int d\ccb\int d\wwb\,\fab(\mathbf{r}_i,\cca,\wwa;\mathbf{r}_j,\ccb,\wwb;t).
\label{III.1}
\eeq
The marginal distribution functions associated with the translational and rotational degrees of freedom are
\beq
\fat(\mathbf{r}_i,\cca;t)=\int d\wwa\,\fa(\mathbf{r}_i,\cca,\wwa;t),\quad \far(\mathbf{r}_i,\wwa;t)=\int d\cca\,\fa(\mathbf{r}_i,\cca,\wwa;t).
\label{III.10}
\eeq
Given a one-body  function $\psi(\cca,\wwa)$, we define its average as
\beq
\langle \psi(\cca,\wwa)\rangle\equiv \frac{1}{\na}\int d\cca \int d\wwa\, \psi(\cca,\wwa) \fa(\cca,\wwa), \quad \na=\int d\cca \int d\wwa\,  \fa(\cca,\wwa),
\label{III.3}
\eeq
where $\na$ is the number density of component $\al$ and, for the sake of brevity, we have omitted the spatial and temporal arguments.
In particular, one can define \emph{partial} temperatures associated with the translational and rotational degrees of freedom as
\beq
\Tat=\frac{\ma}{3}\langle (\cca-\mathbf{u})^2\rangle,\quad \Tar=\frac{\Ia}{3}\langle \wa^2\rangle,
\label{III.11}
\eeq
where
\beq
\mathbf{u}=\frac{\sum_\al \ma\na\langle \cca\rangle}{\sum_\al \ma\na}
\label{III.12}
\eeq
is the flow velocity. Note that in the definition of $\Tar$ we have not referred the angular velocities to any average value because of the lack of invariance under the addition of a common vector to every angular velocity.
The \emph{global} temperature is
\beq
T=\sum_\al\frac{\na}{2n}\left(\Tat+\Tar\right),
\label{III.13}
\eeq
where $n=\sum_\al \na$ is the total number density.

By starting from the Liouville equation and following standard steps, one can derive the Bogoliubov--Born--Green--Kirkwood--Yvon (BBGKY) hierarchy.\cite{BDS97} \ The first equation of the hierarchy reads
\beq
\partial_t \fa(\mathbf{r}_i,\cca,\wwa;t)+\cca\cdot\nabla \fa (\mathbf{r}_i,\cca,\wwa;t)=\sum_\be \Qab[\mathbf{r}_i,\cca,\wwa;t|\fab],
\label{2}
\eeq
where
\beqa
\Qab[\mathbf{r}_i,\cca,\wwa;t|\fab]&=&\dab^2\int d\ccb\int d\wwb\int d\kk\, \Theta(\gh\cdot\kk)(\gh\cdot\kk)\nn
&&\x\Bigg[\frac{1}{\een^2\eet^2}\fab(\mathbf{r}_i,\cca'',\wwa'';\mathbf{r}_i-\dab\kk,\ccb'',\wwb'';t)\nn
&&-
\fab(\mathbf{r}_i,\cca,\wwa;\mathbf{r}_i+\dab\kk,\ccb,\wwb;t)\Bigg]
\label{III.2}
\eeqa
is the collision operator. Here, $\dab\equiv (\da+\db)/2$ and use has been made of Eqs.\ \eqref{restitution} and \eqref{Jacob}.

Multiplying both sides of Eq.\ \eqref{2} by $\psi(\cca,\wwa)$ and integrating over $\cca$ and $\wwa$ one gets
\beq
\partial_t \na \langle \psi(\cca,\wwa)\rangle+\nabla\cdot \na \langle \cca\psi(\cca,\wwa)\rangle=\sum_\be \Iab[\psi(\cca,\wwa)|\fab],
\label{III.5}
\eeq
where
\beqa
\Iab[\psi(\cca,\wwa)|\fab]&\equiv&\int d\cca\int d\wwa\, \psi(\cca,\wwa) \Qab[\cca,\wwa|\fab]\nn
&=&\dab^2\int d\cca\int d\wwa \int d\ccb\int d\wwb\int d\kk\, \Theta(\gh\cdot\kk)(\gh\cdot\kk)
\nn&&\x
\fab(\mathbf{r}_\al,\cca,\wwa;\mathbf{r}_i+\dab\kk,\ccb,\wwb)\left[\psi(\cca',\wwa')-\psi(\cca,\wwa)\right].\nn
\label{3}
\eeqa
Thus, $\na^{-1}\Iab[\psi(\cca,\wwa)|\fab]$ is the \emph{rate of change} of the quantity $\psi(\cca,\wwa)$ due to collisions with particles of component $\be$. This rate of change is a functional of the two-body distribution function $\fab$, as indicated by the notation, and it is in general a rather intricate quantity.
The physically important cases are $\psi(\cca,\wwa)=\{\ma \cca,\Ia \wwa, \ma \ca^2,\Ia \wa^2\}$. The corresponding rates of change are obtained by inserting Eqs.\  \eqref{15b}--\eqref{Z1} into Eq.\ \eqref{3}. Note that so far all the results are formally exact.

To proceed, let us make the approximation
\beq
\Iab[\psi(\cca,\wwa)|\fab]\approx  \Iab[\psi(\cca,\wwa)|\ffab],
\label{III.6}
\eeq
where
\beq
\ffab(\mathbf{r}_\al,\cca,\wwa;\ccb,\wwb)\equiv
\frac{\int d\kk\, \Theta(\gh\cdot\kk)(\gh\cdot\kk)\fab(\mathbf{r}_\al,\cca,\wwa;\mathbf{r}_i+\dab\kk,\ccb,\wwb)}{\int d\kk\, \Theta(\gh\cdot\kk)(\gh\cdot\kk)}
\label{III.7}
\eeq
is the orientational average  of the \emph{pre-collisional} distribution $\fab$. Thus,
Eq.\ \eqref{III.6} replaces a detailed functional of $\fab$ by a simpler one where the solid angle integral
\beq
\int d\kk\, \Theta(\gh\cdot\kk)(\gh\cdot\kk)
\left[\psi(\cca',\wwa')-\psi(\cca,\wwa)\right]
\label{III.8}
\eeq
can be evaluated independently of $\fab$.
It is important to borne in mind  that the approximation \eqref{III.6} is much weaker than the approximation $\fab\approx\ffab$.
Notwithstanding this, the equality $\fab=\ffab$ holds if (a) the system is homogeneous and isotropic (regardless of the reduced densities $\na\da^3$ and $\nb\db^3$), in which case $\fab$ only depends on $|\mathbf{r}_\al-\mathbf{r}_\be|$, or
(b) the system is in the Boltzmann limit ($\na\da^3\to 0$, $\nb\db^3\to 0$), in which case one can formally take $\dab\to 0$ in the contact value of $\fab$. Therefore, the approximation \eqref{III.6} is justified if the density of the granular gas and/or its inhomogeneities are small enough so the value of $\fab$ at contact is hardly dependent on the relative orientation of the two colliding spheres.

In the remainder of this Section we particularize to $\psi(\cca,\wwa)=\{\ma\cca, \Ia\wwa, \ma\ca^2, \Ia\wa^2\}$ and express the rates of change $\na^{-1}\Iab[\psi(\cca,\wwa)]$ in terms of two-body averages of the form
\beqa
\langle A(\cca,\wwa;\ccb,\wwb)\rangle&\equiv &\frac{1}{\na \nb}\int d\cca\int d\wwa \int d\ccb\int d\wwb\, A(\cca,\wwa;\ccb,\wwb)\nn
&&\times\ffab(\cca,\wwa;\ccb,\wwb).
\label{17}
\eeqa
The results are (see the Appendix)
\beq
\na^{-1}\Iab[\ma\cca]=-\nb{\dab^2}\pi\left(\frac{\en+\et}{2}\langle \g\gh\rangle-\frac{2\et}{3}\langle \gh\x\SSab\rangle\right),
\label{16}
\eeq
\beq
\na^{-1}\Iab[\Ia\wwa]=-\nb\dab^2\da\frac{\pi}{8}\et\left[3\langle \g\SSab\rangle-\langle \g^{-1}(\gh\cdot\SSab)\gh\rangle\right],
\label{III.9}
\eeq
\beqa
\na^{-1}\Iab[\ma\ca^2]&=&-\nb{\dab^2}\pi\left[\left({\en+\et}\right)\langle \g\cca\cdot\gh\rangle+\frac{4\et}{3}\langle \SSab\cdot(\cca\x\ccb)\rangle\right.\nn
&&\left.-
\frac{\en^2+\et^2}{2\ma}\langle\g^3\rangle
-\frac{3\et^2}{4\ma}\langle\g\Sab^2\rangle+\frac{\et^2}{4\ma}\langle\g^{-1} \left(\gh\cdot\SSab\right)^2\rangle
\right],
\label{23}
\eeqa
\beqa
\na^{-1}\Iab[\Ia\wa^2]&=&-\nb\dab^2\frac{\pi}{4}\et\Bigg \{{3}\da\langle \g\wwa\cdot\SSab\rangle-\da\langle \g^{-1}\left(\gh\cdot\SSab\right)\left(\gh\cdot\wwa\right)\rangle\nn
&& -\frac{\et}{\ma\qa}\left[2\langle
\g^3\rangle+3\langle\g\Sab^2\rangle-\langle\g^{-1}\left(\gh\cdot\SSab\right)^2\rangle\right]\Bigg\}.
\label{34}
\eeqa

\section{Estimates of the average values}
\label{sec4}
Equations \eqref{16}--\eqref{34} express the collisional rates of change of the main quantities as linear combinations of two-body averages of the form \eqref{17}. They are local functions of space and time and functionals of the orientation-averaged  pre-collisional distribution $\ffab$. While, thanks to the approximation \eqref{III.6}, Eqs.\ \eqref{16}--\eqref{34} are much more explicit than the exact results obtained from Eq.\ \eqref{3}, they still require the full knowledge of  $\ffab$.

\begin{table}[tbp]
\caption{Estimates of the two-body averages appearing in Eqs.\ \protect\eqref{16}--\protect\eqref{34}, as obtained from the replacement \protect\eqref{IV.1}.}
\label{table:1}
\begin{center}
\begin{tabular}{cc} \hline \hline
Quantity & Estimate \\ \hline
$\langle \g\gh\rangle$&$\mathbf{0}$\\
$\langle \gh\x\SSab\rangle$&$\mathbf{0}$\\
$\langle \g\SSab\rangle$&$ \frac{1}{2}\left(\da \bm{\Omega}_\al+\db \bm{\Omega}_\be\right)\langle \g\rangle$\\
$\langle\g^{-1}(\gh\cdot\SSab)\gh\rangle$&$\frac{1}{6}\left(\da \bm{\Omega}_\al+\db \bm{\Omega}_\be\right)\langle \g\rangle$\\
$\langle \g\cca\cdot\gh\rangle$&$\frac{\Tat}{\ma}\left(\frac{\Tat}{\ma}+\frac{\Tbt}{\mb}\right)^{-1}
\langle\g^3\rangle$\\
$\langle \SSab\cdot(\cca\x\ccb)\rangle$&$0$\\
$\langle\g\Sab^2\rangle$&$\left(\frac{3\Tar}{\ma\qa}+\frac{3\Tbr}{\mb\qb}+\frac{1}{2}\da\db \bm{\Omega}_\al\cdot \bm{\Omega}_\be\right)\langle\g\rangle$\\
$\langle\g^{-1} \left(\gh\cdot\SSab\right)^2\rangle$&$\left(\frac{\Tar}{\ma\qa}+\frac{\Tbr}{\mb\qb}+\frac{1}{6}\da\db \bm{\Omega}_\al\cdot \bm{\Omega}_\be\right)\langle\g\rangle$\\
$\langle \g\wwa\cdot\SSab\rangle$&$\left(\frac{6\Tar}{\ma\qa\da}+\frac{1}{2}\db \bm{\Omega}_\al\cdot \bm{\Omega}_\be\right)\langle \g\rangle$\\
$\langle \g^{-1}\left(\gh\cdot\SSab\right)\left(\gh\cdot\wwa\right)\rangle$&$\left(\frac{2\Tar}{\ma\qa\da}+\frac{1}{6}\db \bm{\Omega}_\al\cdot \bm{\Omega}_\be\right)\langle \g\rangle$\\
$\langle\g\rangle$&$2\sqrt{\frac{2}{\pi}}\chiab\left(\frac{\Tat}{\ma}+\frac{\Tbt}{\mb}\right)^{1/2}$\\
$\langle\g^3\rangle$&$8\sqrt{\frac{2}{\pi}}\chiab\left(\frac{\Tat}{\ma}+\frac{\Tbt}{\mb}\right)^{3/2}$\\
\hline
\end{tabular}
\end{center}
\end{table}

Suppose, for simplicity, that $\langle \cca\rangle=\langle \ccb\rangle=\mathbf{u}$ and define the average angular velocities
\beq
\langle \wwa\rangle=\bm{\Omega}_\al,\quad \langle \wwb\rangle=\bm{\Omega}_\be.
\label{[IV.0}
\eeq
Now, let us imagine that, instead of the full knowledge of $\ffab$, we only know the local values of the two densities ($\na$ and $\nb$), the two average angular velocities ($\bm{\Omega}_\al$ and $\bm{\Omega}_\be$), and the four partial temperatures ($\Tat$, $\Tar$, $\Tbt$, and $\Tbr$). The question we want to address in this section is, can we get reasonable \emph{estimates} of the two-body averages appearing in Eqs.\ \eqref{16}--\eqref{34} by expressing them in terms of  $\na$, $\nb$, $\bm{\Omega}_\al$, $\bm{\Omega}_\be$, $\Tat$, $\Tar$, $\Tbt$, and $\Tbr$?
In the absence of further information, the least biased estimates are obtained from the replacement
\beqa
\ffab(\cca,\wwa;\ccb,\wwb)&\to& \chiab \left(\frac{\ma\mb}{4\pi^2\Tat\Tbt}\right)^{3/2}\exp\left[-\ma\frac{(\cca-\mathbf{u})^2}{2\Tat}-\mb\frac{(\ccb-\mathbf{u})^2}{2\Tbt}\right]
\nn
&&\times\far(\wwa)\fbr(\wwb),
\label{IV.1}
\eeqa
where $\chiab$ is the contact value of the pair correlation function. Equation \eqref{IV.1} can be justified by maximum-entropy arguments, except that here we do not need to assume a Maxwellian form for the rotational distributions, given that the angular velocities only appear linearly or quadratically in  Eqs.\ \eqref{16}--\eqref{34}. Like in the approximation \eqref{III.6}, it is important to stress that we are not making the strong claim that $\ffab(\cca,\wwa;\ccb,\wwb)$ is well approximated  by the right-hand side of Eq.\ \eqref{IV.1} [see Ref.\ \citen{BPKZ07}] but only the wekaer one that the two-body averages can be estimated by performing such a replacement. Table \ref{table:1} displays those estimates.

Although $\langle \cca\rangle=\langle \ccb\rangle$ has been assumed in the results shown in Table \ref{table:1}, the generalization to $\langle \cca\rangle\neq\langle \ccb\rangle$ can be carried out following the same steps as  done in Ref.\  \citen{VGS07} for smooth spheres.

\section{Energy production  rates and cooling rate}
\label{sec5}
The most characteristic feature of a granular gas is the energy dissipation taking place after each collision. In the model of inelastic rough hard spheres this is clearly apparent from Eq.\ \eqref{29}. On the other hand, any of the four partial kinetic energies contributing to $E_{\al\be}$ in Eq.\ \eqref{Z2} can either increase or decrease after a given collision. To characterize this effect at a statistical level, it is convenient to introduce the rates of change of the partial temperatures $\Tat$ and $\Tar$ due to collisions of particles of component $\al$ with particles of component $\be$. More explicitly, we define the (partial) energy production rates $\zabt$ and $\zabr$ as
\beq
\zabt\equiv -\frac{1}{3\na\Tat}\Iab[\ma(\cca-\mathbf{u})^2],\quad \zabr\equiv -\frac{1}{3\na\Tar}\Iab[\Ia\wa^2].
\label{54}
\eeq
When collisions of particles of component $\al$ with all the components  are considered, we get the (total) energy production rates
\beq
\xi^\tr_\al\equiv-\frac{1}{\Tat}\left(\frac{\partial\Tat}{\partial t}\right)_{\text{coll}}=\sum_\be \zabt,\quad
\xi^\rot_\al\equiv-\frac{1}{\Tar}\left(\frac{\partial\Tar}{\partial t}\right)_{\text{coll}}=\sum_\be \zabr.
\label{107}
\eeq
Finally, the net \emph{cooling} rate is
\beq
\zeta\equiv-\frac{1}{T}\left(\frac{\partial T}{\partial t}\right)_{\text{coll}}=\sum_\al \frac{\na}{2nT}
\left(\Tat\xi^\tr_\al+\Tar\xi^\rot_\al\right).
\label{110}
\eeq
In contrast to the energy production rates defined in Eqs.\ \eqref{54} and \eqref{107}, the cooling rate $\zeta$ is positive definite, i.e., collisions produce a decrease of the total temperature $T$ unless $\een=1$ and $\eet=\pm 1$ for \emph{all} pairs $\al\be$.

{}From Eqs.\ \eqref{16}--\eqref{34} and the expressions of Table \ref{table:1} one gets the following \emph{estimates}:
\beq
\na^{-1}\Iab[\ma\cca]=\mathbf{0},\quad \na^{-1}\Iab[\Ia\wwa]=-{\nu_{\al\be}}\frac{\et}{4}\da\left(\da\bm{\Omega}_\al+\db\bm{\Omega}_\be\right),
\label{V.1}
\eeq
\beqa
\zabt&=&\frac{\nu_{\al\be}}{\ma\Tat}
\left[2\left({\en+\et}\right){\Tat}-
\left({\en^2+\et^2}\right)\left(\frac{\Tat}{\ma}+\frac{\Tbt}{\mb}\right)\right.\nn
&&\left.
-{\et^2}\left(\frac{\Tar}{\ma\qa}+\frac{\Tbr}{\mb\qb}+\frac{1}{6}\da\db \bm{\Omega}_\al\cdot \bm{\Omega}_\be\right)
\right],
\label{55}
\eeqa
\beqa
\zabr&=&\frac{\nu_{\al\be}}{\ma\qa\Tar}\et\left[2{\Tar}+\frac{1}{6}\ma\qa\da\db \bm{\Omega}_\al\cdot \bm{\Omega}_\be-{\et}\left(\frac{\Tat}{\ma}+\frac{\Tbt}{\mb}+\frac{\Tar}{\ma\qa}+\frac{\Tbr}{\mb\qb}\right.\right.\nn
&&\left.\left.+\frac{1}{6}\da\db \bm{\Omega}_\al\cdot \bm{\Omega}_\be\right)
\right],
\label{56}
\eeqa
where we have introduced the effective collision frequency
\beq
\nu_{\al\be}\equiv\frac{4\sqrt{2\pi}}{3}\chiab\nb{\dab^2}\sqrt{\frac{\Tat}{\ma}+\frac{\Tbt}{\mb}}.
\label{56b}
\eeq

Equations \eqref{55} and \eqref{56} are the main results of this paper. They express the collisional rates of change as functions of the local values of  $\na$, $\nb$, $\bm{\Omega}_\al$, $\bm{\Omega}_\be$, $\Tat$, $\Tar$, $\Tbt$, and $\Tbr$, as well as of the mechanical parameters $\ma$, $\mb$, $\da$, $\db$, $\qa$, $\qb$, $\een$, and $\eet$. The energy production rates \eqref{55} and \eqref{56} can be decomposed into two classes of  terms. The first class is made of terms  headed by $1+\een$ and $1+\eet$ which exist even if collisions are conservative. These terms do not have a definite sign, are proportional to temperature differences (except for the scalar product $\bm{\Omega}_\al\cdot \bm{\Omega}_\be$), and tend to make the four temperatures equal . The second class is made of terms headed by ${1-\een^2}$ and $1-\eet^2$ and are positive definite, thus contributing to a decrease of the temperatures due to energy dissipation. The terms of the first and second classes can be termed \emph{equipartition} and \emph{cooling} rates, respectively. Only the latter class contributes to the net cooling rate defined by Eq.\ \eqref{110}. The result is
\beqa
\zeta&=&\sum_{\al\be} \frac{\na\nu_{\al\be}}{4nT}\mab\left[(1-\een^2)\left(\frac{\Tat}{\ma}+\frac{\Tbt}{\mb}\right)+\frac{\qab}{1+\qab}
(1-\eet^2)
\left(\frac{\Tat}{\ma}+\frac{\Tbt}{\mb}\right.\right.\nn
&&\left.\left.+\frac{\Tar}{\ma\qa}+\frac{\Tbr}{\mb\qb}+\frac{1}{6}\da\db \bm{\Omega}_\al\cdot \bm{\Omega}_\be\right)\right].
\label{114}
\eeqa
In the case of smooth spheres ($\eet=-1$), Eqs.\ \eqref{55} and \eqref{114} reduce to those obtained in Ref.\ \citen{GD99}.

Before closing this section, note that in the monodisperse case  Eqs.\  \eqref{55}, \eqref{56}, and \eqref{114} become
\beq
\zt=\frac{\nu}{2}\left[1-\esn^2+\frac{\q}{1+\q}\left(1-\est^2\right)+\frac{\q}{(1+\q)^2}\left(1+\est\right)^2\left(1-\frac{\Tr+\q m\ds^2\Omega^2/12}{\Tt}\right)\right],
\label{61}
\eeq
\beqa
\zr&=&\frac{\nu}{2}\frac{1+\est}{1+\q}\frac{\Tt}{\Tr}\left[(1-\est)\frac{\Tr+\q m\ds^2\Omega^2/12}{\Tt}-\frac{\q}{1+\q}\left(1+\est\right)\right.\nn
&&\left.\times\left(1-\frac{\Tr+\q m\ds^2\Omega^2/12}{\Tt}\right)\right],
\label{62}
\eeqa
\beq
\zeta=\frac{\zt \Tt+\zr \Tr}{\Tt+\Tr}=\frac{\nu}{2}\frac{\Tt}{\Tt+\Tr}\left[1-\esn^2+\frac{1-\est^2}{1+\q}\left(\q+\frac{\Tr+\q m\ds^2\Omega^2/12}{\Tt}\right)\right],
\label{63}
\eeq
where $\nu\equiv (4\sqrt{2\pi}/3)\chi n\ds^2\sqrt{2\Tt/m}$. Equations \eqref{61}--\eqref{63} agree with those previously derived in Refs.\ \citen{GS95,LHMZ98} with $\bm{\Omega}=\mathbf{0}$.
\section{Application to the homogeneous cooling state}
\label{sec6}
In the so-called homogeneous cooling state (HCS) the flux term $\nabla\cdot \na \langle \cca\psi(\cca,\wwa)\rangle$ in Eq.\ \eqref{III.5} is absent. Therefore, the evolution equations for the total and partial temperatures are
\beq
 \partial_t T=-\zeta T,\quad \partial_t\frac{\Tat}{T}=-\left(\zt_\al-\zeta\right)\frac{\Tat}{T},\quad
\partial_t\frac{\Tar}{T}=-\left(\zr_\al-\zeta\right)\frac{\Tar}{T}.
\label{57}
\eeq
After a certain transient time, a scaling regime is reached where all the time dependence occurs through the total temperature $T$, which implies constant temperature ratios and equal production rates, i.e.,
\beq
\zt_1=\zt_2=\cdots=\zr_1=\zr_2=\cdots.
\label{59}
\eeq
Since the HCS is an isotropic state, it follows that, for symmetry, $\bm{\Omega}_\al=\mathbf{0}$.

\subsection{Monodisperse system}

In the monodisperse case the HCS condition $\zt=\zr$ yields the quadratic equation
\beq
1-\esn^2-\frac{1-\q}{1+\q}(1-\est^2)-\frac{\q}{(1+\q)^2}(1+\est)^2\left(\frac{\Tr}{\Tt}-\frac{\Tt}{\Tr}\right)=0.
\label{79}
\eeq
In the smooth-sphere limit ($\est \to -1$) we get
\beq
\frac{\Tr}{\Tt}\approx
\begin{cases}
\frac{(1+\q)^2}{\q}(1-\esn^2)(1+\est)^{-2}\to \infty,&\esn<1,\\
\frac{\q}{2(1-\q^2)}(1+\est)\to 0,&\esn=1.
\end{cases}
\label{80}
\eeq
Thus the elastic-sphere limit ($\esn\to 1$) and the smooth-sphere limit ($\est\to -1$) do not commute. If the spheres are inelastic ($\esn<1$)  but perfectly smooth ($\est= -1$), the rotational and translational degrees of freedom are decoupled  and $\Tr$ does not change with time, while $\Tt$ keeps decreasing due to inelasticity \cite{HHZ00}. As a consequence, the ratio $\Tr/\Tt$ diverges in the long-time limit. On the other hand, if first we assume that the spheres are perfectly elastic ($\esn=1$) and then consider small roughness ($\est\to -1$), the coupling between $\Tr$ and $\Tt$ is weak but not broken. As long as $\Tt\sim\Tr$, the translational temperature decays more slowly than the rotational temperature ($\zt/\zr\approx\q$), resulting eventually in a temperature ratio $\Tr/\Tt\sim 1+\est\to 0$.

\subsection{Binary mixture}
In the particular case of a binary mixture, there exist three relevant temperature ratios that can be chosen in different ways. Here we take one of the translational/rotational ratios ($\Tt_1/\Tr_1$) and the two component/component ratios ($\Tt_2/\Tt_1$ and $\Tr_2/\Tr_1$). The dimensionless parameter space is twelve-dimensional: the three coefficients of normal restitution ($\esn_{11}$, $\esn_{12}$, $\esn_{22}$), the three coefficients of tangential restitution ($\est_{11}$, $\est_{12}$, $\est_{22}$), the two parameters $\q_1$ and $\q_2$, the mass ratio $m_1/m_2$, the size ratio $\ds_1/\ds_2$, the mole fraction $x_1=n_1/(n_1+n_2)$, and the total packing fraction $\phi=(\pi/6)(n_1\ds_1^3+n_2\ds_2^3)$.

\begin{figure}[tbp]
\centerline{\includegraphics[width=\columnwidth]{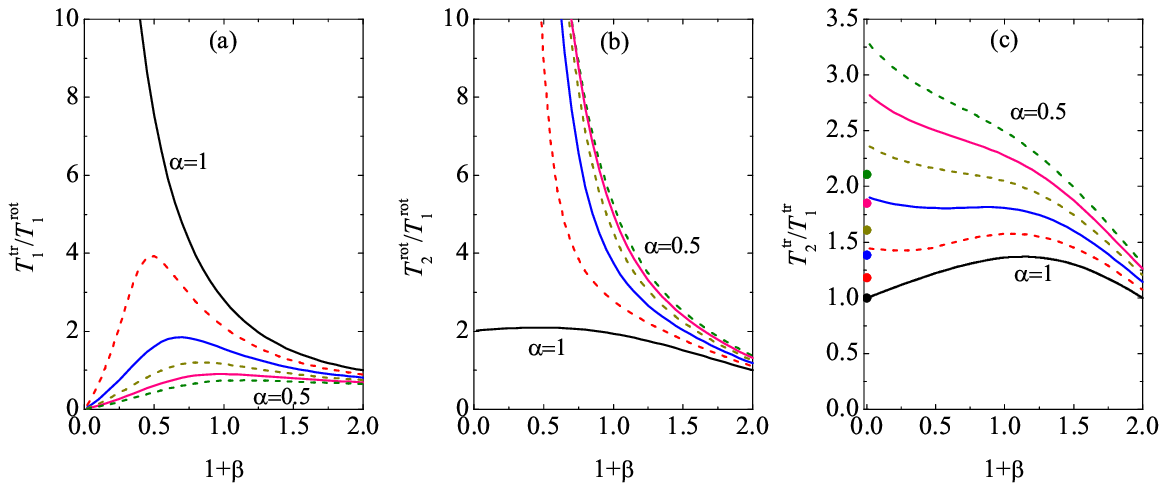}}
\caption{HCS of a dilute equimolar binary mixture with $\esn_{11}=\esn_{12}=\esn_{22}=\esn$, $\est_{11}=\est_{12}=\est_{22}=\est$, $\q_1=\q_2=\frac{2}{5}$, $\ds_2/\ds_1=2$, and $m_2/m_1=8$. Plot of the temperature ratios (a) $\Tt_1/\Tr_1$, (b) $\Tr_2/\Tr_1$, and (c) $\Tt_2/\Tt_1$ vs $1+\est$ for [from top to bottom in (a) and from bottom to top in (b) and (c)] $\esn=1$, $0.9$, $0.8$, $0.7$, $0.6$, and $0.5$. The circles at $1+\est=0$ in (c) represent the results obtained in the case of perfectly smooth spheres (rotational degrees of freedom ignored from the beginning).\protect\cite{GD99}}
\label{fig2}
\end{figure}

To illustrate the results, here we first assume an equimolar mixture where all the spheres are uniformly solid  and are made of the same material, the size of the spheres of one component being twice that of  the other component. More specifically, $x_1=\frac{1}{2}$, $\esn_{11}=\esn_{12}=\esn_{22}=\esn$, $\est_{11}=\est_{12}=\est_{22}=\est$, $\q_1=\q_2=\frac{2}{5}$, $\ds_2/\ds_1=2$, and $m_2/m_1=8$. Moreover, we consider a dilute granular gas ($\phi\ll1$), so that $\chi_{\al\be}\approx 1$. Thus only the parameters $\esn$ and $\est$ remain free.
Figure \ref{fig2} shows the three independent temperature ratios as functions of the roughness parameter $1+\est$ for several characteristic values of the inelasticity parameter $\esn$. As happened in the monodisperse case, the translational/rotational temperature ratio $\Tt_1/\Tr_1$ exhibits a peculiar behavior in the smooth-sphere limit  $1+\est\to 0$: it diverges for elastic particles ($\esn=1$) while it vanishes for inelastic particles ($\esn<1$). This phenomenon has a reflection in the rotational/rotational ratio: either $\Tr_2/\Tr_1$ converges to a finite value or it diverges, depending on whether $\esn=1$ or $\esn<1$, respectively. Quite interestingly, the huge disparity between the rotational and translational temperatures in the smooth-sphere limit has a non-negligible effect on the translational/translational ratio $\Tt_2/\Tt_1$ if $\esn<1$:  it tends to a finite value different from (in fact higher than) the value directly obtained in the case of perfectly smooth spheres.\cite{GD99}
Thus, a tiny amount of roughness has  dramatic effects on the temperature ratio $\Tt_2/\Tt_1$, producing an enhancement of non-equipartition.

\begin{figure}[tbp]
\centerline{\includegraphics[width=\columnwidth]{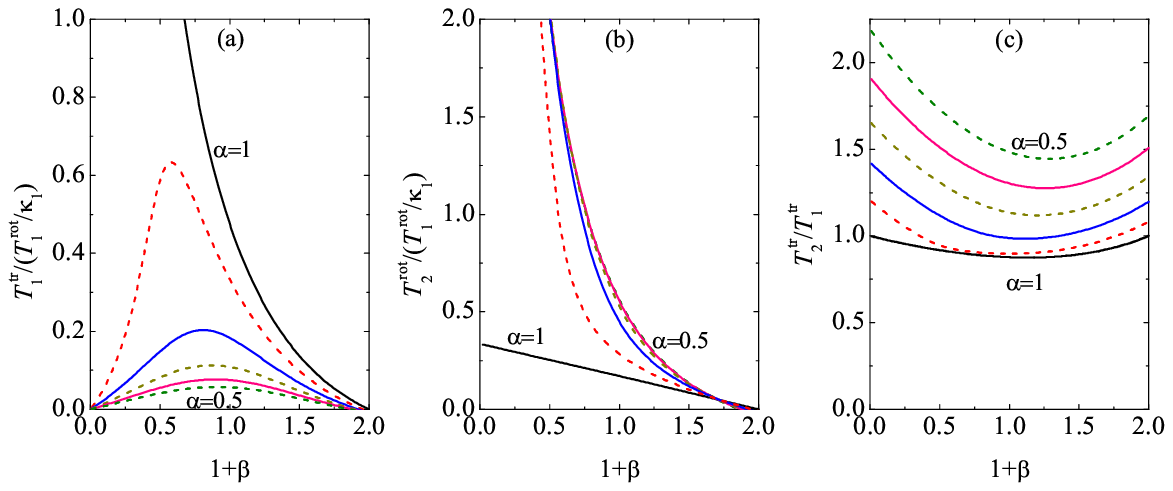}}
\caption{HCS of a dilute equimolar binary mixture with $\esn_{11}=\esn_{12}=\esn_{22}=\esn$, $\est_{11}=\est_{12}=\est_{22}=\est$, $\q_1\to 0$, $\q_2=\frac{2}{3}$, $\ds_2/\ds_1=1$, and $m_2/m_1=1$. Plot of the temperature ratios (a) $\Tt_1/(\Tr_1/\q_1)$, (b) $\Tr_2/(\Tr_1/\q_1)$, and (c) $\Tt_2/\Tt_1)$ vs $1+\est$ for [from top to bottom in (a) and from bottom to top in (b) and (c)] $\esn=1$, $0.9$, $0.8$, $0.7$, $0.6$, and $0.5$.}
\label{fig3}
\end{figure}

As a second example, we now consider a dilute equimolar mixture of spheres externally identical (same mass, size, and coefficients of restitution), except that the mass of the spheres in one of the components is practically concentrated in the centre ($\q_1\to 0$), while the other component is made of hollow spheres ($\q_2=\frac{2}{3}$). Were the spheres \emph{strictly} smooth, then the system would be indistinguishable from a monodisperse system. However, again the smooth case is singular and the results show that both components have different temperatures even in the limit $1+\est\to 0$, as illustrated by Fig.\ \ref{fig3}. Note that in the limit $\q_1\to 0$ the rotational temperature $\Tr_1$ vanishes but the ratio $\Tr_1/\q_1$ is well defined.

\subsection{Locus of equipartition}

\begin{figure}[tbp]
\centerline{\includegraphics[width=\columnwidth]{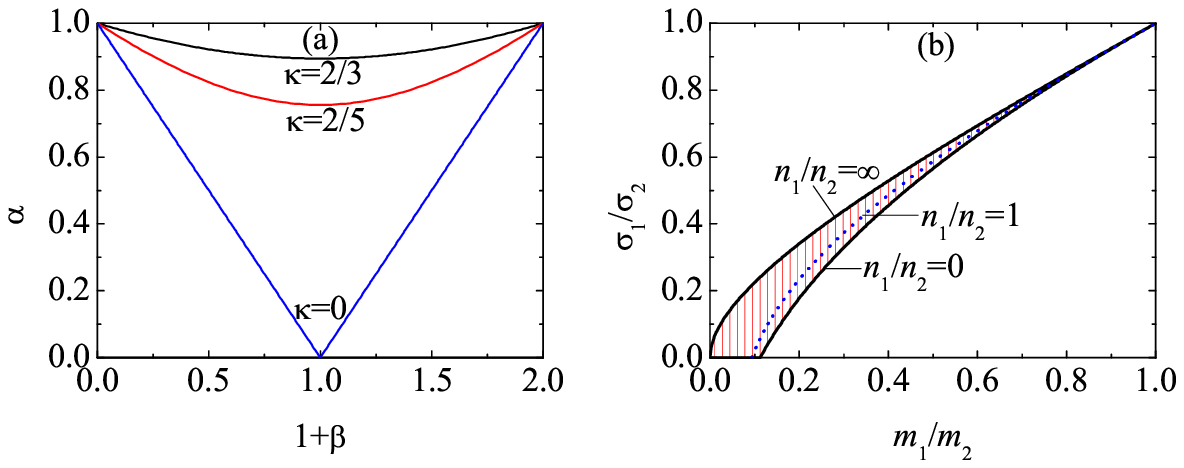}}
\caption{HCS of a dilute  binary mixture with $\esn_{11}=\esn_{12}=\esn_{22}=\esn$, $\est_{11}=\est_{12}=\est_{22}=\est$, and $\q_1=\q_2=\q$. (a) Loci of values of $\esn$ and $\est$ for $\q=\frac{2}{3}$, $\frac{2}{5}$, and 0 where equipartition holds, provided the second equality of Eq.\ \protect\eqref{115} is satisfied. (b) The hatched region represents the values of $m_1/m_2$ and and $\ds_1/\ds_2$ where equipartition is possible. The boundaries of the region correspond to the extreme limits $n_1/n_2\to 0$ and $n_1/n_2\to \infty$, while the dotted line corresponds to $n_1/n_2=1$.}
\label{fig4}
\end{figure}

As is well known, non-equipartition prevails in granular mixtures. This has been illustrated in Figs.\ \ref{fig2} and \ref{fig3}. On the other hand, by fine-tuning the mechanical parameters and the composition, it is in principle possible to reach equipartition, i.e., $\Tt_\al=\Tr_\al=T$. In this equipartition case, Eqs.\ \eqref{55} and \eqref{56} become
\beq
\zabt={\nu_{\al\be}}\frac{\mb}{\ma+\mb}\left[1-\een^2+\frac{\qab}{1+\qab}\left(1-\eet^2\right)\right],\quad \zabr={\nu_{\al\be}}\frac{\mb\qb}{\ma\qa+\mb\qb}\frac{1-\eet^2}{1+\qab},
\label{104b}
\eeq
where $\nu_{\al\be}=\frac{4\sqrt{2\pi}}{3}\chi_{\al\be}\nb{\dab^2}\sqrt{{T(\ma+\mb)}/{\ma\mb}}$.  To show under which conditions equipartition is possible, let us  consider again a dilute binary mixture with $\esn_{11}=\esn_{12}=\esn_{22}=\esn$, $\est_{11}=\est_{12}=\est_{22}=\est$, and $\q_1=\q_2=\q$. Otherwise,  $\esn$, $\est$, $\q$, $m_1/m_2$, $\ds_1/\ds_2$, and $n_2/n_1$ are arbitrary.
After simple algebra, the HCS conditions \eqref{59} yield
\beq
1-\esn^2=\frac{1-\q}{1+\q}(1-\est^2),\quad \frac{n_1}{n_2}=\frac{\ds_{12}^2\sqrt{\frac{m_2}{m_1}}-\ds_2^2\sqrt{\frac{m_1+m_2}{2m_2}}}{\ds_{12}^2\sqrt{\frac{m_1}{m_2}}-\ds_1^2\sqrt{\frac{m_1+m_2}{2m_1}}}.
\label{115}
\eeq
The first equality establishes a relationship between both coefficients of restitution and $\q$ that is independent of composition, masses, and sizes of the particles [see Fig.\ \ref{fig4}(a)]. Moreover, the composition is constrained by  the second equality.  Without loss of generality, let us take $m_1\leq m_2$. Then, positivity of the right-hand side of the second equality of Eq.\ \eqref{115} implies that
$\text{max}\left\{0,A_{12}\right\}\leq{\ds_1}/{\ds_2}\leq A_{21}^{-1}\leq 1$,
where   $A_{\al\be}\equiv 2\left[\ma(\ma+\mb)/2\mb^2\right]^{1/4}-1$.
The above inequality defines a wing-shaped region in the plane $\ds_1/\ds_2$ vs $m_1/m_2$ [see Fig.\ \ref{fig4}(b)] where equipartition is possible, provided both equalities in Eq.\ \eqref{115} are satisfied.

\section{Concluding remarks}
\label{sec7}

Despite the success and prevalence of the inelastic smooth-sphere model to characterize granular matter under rapid flow conditions,\cite{G03} it is known that the presence of friction in collisions can be relevant and even essential for a more realistic modeling of granular flows.\cite{MSS04,BPKZ07} Moreover, granular matter is typically present in nature in polydisperse form.
In this paper we have combined both aspects (roughness plus polydispersity) in a model of inelastic hard spheres with constant coefficients of normal ($\een$) and tangential ($\eet$) restitution. Otherwise, the masses, diameters, moments of inertia, coefficients of restitution, and composition are arbitrary. Given the complexity of the system, here we focus on the derivation of the (partial) energy production rates $\zabt$ and $\zabr$ due to binary collisions. These quantities determine the collisional rates of change of the translational ($\Tat$) and rotational ($\Tar$) granular temperatures  associated with each component, as well as the cooling rate $\zeta$ of the total temperature of the mixture. They can be considered as the most basic quantities in a granular mixture since they are involved in the energy balance equations.

Starting from the collisional rules \eqref{13b} and \eqref{14}, together with Eq.\ \eqref{15}, the energy production rates are first expressed in a formally exact way in terms of the two-body distribution function $\fab$ [see Eqs.\ \eqref{3}, \eqref{15c}, and \eqref{Z1}]. Next, the true function $\fab$ is replaced by its pre-collisional orientational average $\ffab$ defined by Eq.\ \eqref{III.7}, what is justified if the density and/or the inhomogeneities are not too large. This replacement allows us to express the collisional rates of change as combinations of two-body averages, Eqs.\ \eqref{16}--\eqref{34}. To get more explicit results as functions of the local values of densities, temperatures, and mean angular velocities, a maximum-entropy approach is followed in Eq.\ \eqref{IV.1}. The final expressions are given by Eqs.\ \eqref{55}, \eqref{56}, and \eqref{114}. These are the main results of the paper and generalize previous results derived for monodisperse rough-sphere gases \cite{Z06} and polydisperse smooth-sphere gases.\cite{GD99}

As a preliminary application of our results, we have studied the HCS, where collisions are the only source of energy change. For a binary mixture of common coefficients of restitution $\een=\esn$ and $\eet=\est$, we have analyzed the influence of both inelasticity and roughness on the three independent temperature ratios $T_1^\text{tr}/T_1^\text{rot}$, $T_2^\text{tr}/T_1^\text{tr}$, and $T_2^\text{rot}/T_1^\text{rot}$. As a surprising result, we find that a small amount of roughness has a significant effect on the translational/translational ratio $T_2^\text{tr}/T_1^\text{tr}$. More specifically, it turns out that, at a given value of $\esn$, the value of $T_2^\text{tr}/T_1^\text{tr}$ in the limit $\est\to -1$ differs from the value obtained directly in the smooth case ($\est= -1$). This paradoxical phenomenon is mainly due to the fact that, even if $\est$ is close to $-1$, there exists a transfer of energy from the rotational to the translational degrees of freedom that becomes relevant since the rotational temperatures are much larger than the translational ones in the nearly smooth case. The latter effect is closely tied to the non-stationary character of the HCS. In homogeneous steady states, as the one driven by a white-noise thermostat, the extreme sensitivity of the ratio $T_2^\text{tr}/T_1^\text{tr}$ to whether $\est\to -1$ or $\est= -1$ disappears.

The results obtained here can be applied to several problems. First, we plan to compare the theoretical results derived in this paper with computer simulations in homogeneous states. This will allow us to assess the  reliability of the approximations made here to get explicit results. Secondly, taking the local version of the HCS as the reference state, a Chapman--Enskog method can be followed to get the Navier--Stokes constitutive equations for a mixture of rough spheres. Finally, following  steps similar to those done in Ref.\ \citen{VGS07}, a kinetic model for rough spheres preserving the energy production rates obtained here will be proposed.

\section*{Acknowledgements}
This work was initiated during a two-week visit (between  January and February 2009) of G.M.K. to the Departamento  de F\'{\i}sica, Universidad de Extremadura. He is grateful to this institution for its hospitality  and support.
A.S. acknowledges the hospitality of the Yukawa Institute for Theoretical Physics (Kyoto University, Japan) during the long-term workshop ``Frontiers in Nonequilibrium Physics. Fundamental Theory, Glassy \& Granular Materials, and Computational Physics,'' chaired by Prof.\ Hisao Hayakawa. A.S. is particularly grateful to his roommates J. W. Dufty, I. Goldhirsch, and J. T. Jenkins for insightful discussions.
The research of A.S and V.G. has been supported by the Ministerio de Educaci\'on y Ciencia (Spain) through Grant No.\
FIS2007-60977 (partially financed by FEDER funds) and by the Junta
de Extremadura through Grant No.\ GRU09038.
While this paper was in the proofs stage, Prof.\ Isaac Goldhirsch sadly passed away.  We want to dedicate our paper to his fond memory.

\appendix

\section*{Collisional changes}
\label{appA}

In this Appendix the expressions for the collisional changes of $\cca$, $\wwa$, $\ca^2$, and $\wa^2$ are given.
First, note that
from Eq.\ \eqref{15} it follows that
\beq
\J^2
=\en^2(\gh\cdot\kk)^2+\et^2\left[(\kk\x\gh)^2+(\kk\x\SSab)^2-2\gh\cdot(\kk\x\SSab)\right],
\label{5}
\eeq
\beq
\kk\x\JJ=\et\left[\kk\x\gh-\kk\x(\kk\x\SSab)\right],
\label{27}
\eeq
\beq
\left(\kk\x\JJ\right)^2=\et^2\left[(\kk\x\gh)^2+(\kk\x\SSab)^2+2(\kk\x\gh)\cdot\SSab\right].
\label{28}
\eeq
where use has been made of the mathematical identities
\beq
\kk\x(\kk \x\mathbf{A})=(\kk\cdot \mathbf{A})\kk-\mathbf{A}, \quad (\kk\x\mathbf{A})\cdot(\kk\x\mathbf{B})=\mathbf{A}\cdot\mathbf{B}-(\kk\cdot\mathbf{A})(\kk\cdot\mathbf{B}).
\label{1}
\eeq
Next, Eqs.\ \eqref{13b} and \eqref{14} yield
\beq
\ma \cca'-\ma\cca=-\en (\gh\cdot\kk)\kk-\et\left[\gh-\kk\x\SSab-(\gh\cdot\kk)\kk\right],
\label{15b}
\eeq
\beq
\Ia \wwa'-\Ia\wwa=-\frac{\da}{2}\et\left[\kk\x\gh-\kk\x(\kk\x\SSab)\right],
\label{27b}
\eeq
\beqa
\ma {\ca'}^2-\ma\ca^2&=&\frac{\en^2}{\ma}(\gh\cdot\kk)^2+\frac{\et^2}{\ma}\left[(\kk\x\gh)^2+(\kk\x\SSab)^2-2\gh\cdot(\kk\x\SSab)\right]\nn
&&
-2\en (\gh\cdot\kk)(\cca\cdot\kk)-2\et\left[(\kk\x\cca)\cdot(\kk\x\gh)-\cca\cdot(\kk\x\SSab)\right],\nn
\label{15c}
\eeqa
\beqa
\Ia{\wa'}^2-\Ia\wa^2&=&
\frac{\et^2}{\ma\qa}\left[(\kk\x\gh)^2+(\kk\x\SSab)^2+2(\kk\x\gh)\cdot\SSab\right]
\nn
&&-\et\da\wwa\cdot\left[\kk\x\gh-\kk\x(\kk\x\SSab)\right].
\label{Z1}
\eeqa
Similar expressions are obtained for particle $j$ by exchanging $i\leftrightarrow j$ and $\kk\leftrightarrow -\kk$.
Combining Eqs.\ \eqref{15c} and \eqref{Z1}, plus their counterparts for particle $j$, one can get Eq.\ \eqref{29}, where use is made of Eq.\ \eqref{20} and the identity $\mathbf{A}\cdot(\kk\x\mathbf{B})=-\mathbf{B}\cdot(\kk\x\mathbf{A})$.
{}From Eqs.\ \eqref{15b}--\eqref{Z1} one can easily obtain Eqs.\ \eqref{16}--\eqref{34} by using the mathematical identities
\beq
\int d\kk \Theta(\gh\cdot\kk)(\gh\cdot\kk)^\ell=\frac{2\pi}{\ell+1} \g^\ell,
\label{6}
\eeq
\beq
\int d\kk \Theta(\gh\cdot\kk)(\gh\cdot\kk)^\ell \kk=\frac{2\pi}{\ell+2} \g^{\ell-1}\gh,
\label{7}
\eeq
\beq
\int d\kk \Theta(\gh\cdot\kk)(\gh\cdot\kk)^\ell \kk\kk=\frac{2\pi}{(\ell+1)(\ell+3)} \g^{\ell-2}\left(\ell\gh\gh+\g^2 \mathsf{I}\right),
\label{8}
\eeq
where  $\mathsf{I}$ is the unit tensor.


\begin{thebibliography}{99}

\bibitem{G03}
I. Goldhirsch, \JL{Annu.\ Rev.\ Fluid Mech.,35,2003,267}.

\bibitem{BSSP04}
N. Brilliantov, C. Salue\~na, T. Schwager, and T. P\"oschel, \PRL{93,2004,134301}.

\bibitem{XVKL09}
H. Xu, R. Verberg, D. L. Koch, and M. Y. Louge, \JL{J.\ Fluid Mech.,618,2009,181}.

\bibitem{CZMP09}
R. Cruz Hidalgo, I. Zuriguel, D. Maza, and I. Pagonabarraga, \PRL{103,2009,118001}.

\bibitem{G08}
V. Garz\'o, in \emph{Theory and Simulation of Hard-Sphere Fluids and Related Systems}, edited by A. Mulero (Springer, Berlin, 2008), pp. 493--540.

\bibitem{MSS04}
S. J. Moon, J. B. Swift, and H. L. Swinney, \PRE{69,2004,031301}.

\bibitem{GD99}
V. Garz\'o and J. W. Dufty, \PRE{60,1999,5706}

\bibitem{MG02}
J. M. Montanero and V. Garz\'o, \JL{Gran.\ Matt.,4,2002,17}

\bibitem{BT02}
A. Barrat and E. Trizac, \JL{Gran.\ Matt.,4,2002,57}.

\bibitem{DHGD02}
S. R. Dahl, C. M. Hrenya, V. Garz\'o, and J. W. Dufty \PRE{66,2002,041301}.

\bibitem{JM89}
J. T. Jenkins and F. Mancini, \JL{Phys.\ Fluids A,1,1989,2050}.

\bibitem{GD02}
V. Garz\'o and J. W. Dufty, \JL{Phys.\ Fluids,14,2002,1476}.

\bibitem{SGNT06}
D. Serero, I. Goldhirsch, S. H. Noskowicz, M.-L. Tan, \JL{J.\ Fluid Mech.,554,2006,237}.

\bibitem{GDH07}
V. Garz\'o, J. W. Dufty, and C. M. Hrenya, \PRE{76,2007,031303}.

\bibitem{GHD07}
V. Garz\'o,  C. M. Hrenya, and J. W. Dufty, \PRE{76,2007,031304}.

\bibitem{HQL01}
D. C. Hong, P. V. Quinn, and S. Luding, \PRL{86,2001,3423}.

\bibitem{JY02}
J. T. Jenkins and D. K. Yoon, \PRL{88,2002,194301}.

\bibitem{K04}
A. Kudrolli, \JL{Rep.\ Progr.\ Phys.,67,2004,209}.

\bibitem{BRM05}
J. J. Brey, M. J. Ruiz-Montero, and F. Moreno, \PRL{95,2005,098001}.

\bibitem{G08a}
V. Garz\'o, \PRE{78,2008,020301(R)}.

\bibitem{JR85}
J. T. Jenkins and M. W. Richman, \JL{Phys.\ Fluids,28,1985,3485}.

\bibitem{LS87}
C. K. K. Lun and S. B. Savage, \JL{J.\ Appl.\ Mech.,54,1987,47}.

\bibitem{C89}
C. S. Campbell, \JL{J.\ Fluid Mech.,203,1989,449}.

\bibitem{L91}
C. K. K. Lun, \JL{J.\ Fluid Mech.,233,1991,539}.

\bibitem{LN94}
C. K. K. Lun and A. A. Bent, \JL{J.\ Fluid Mech.,258,1994,335}.

\bibitem{L96}
C. K. K. Lun, \JL{Phys.\ Fluids,8,1996,2868}.

\bibitem{ZVPSH98}
P. Zamankhan, H. V. Tafreshi, W. Polashenski, P. Sarkomaa, and C. L. Hyndman, \JL{J.\ Chem.\ Phys.,109,1998,4487}.

\bibitem{JZ02}
J. T. Jenkins and C. Zhang,  \JL{Phys.\ Fluids,14,2002,1228}.

\bibitem{PZMZ02}
W. Polashenski,  P. Zamankhan, S. M\"akiharju, and P. Zamankhan, \PRE{66,2002,021303}.


\bibitem{GS95}
A. Goldshtein and M. Shapiro, \JL{J.\ Fluid Mech.,282,1995,75}.


\bibitem{HZ98}
M. Huthmann,  and A. Zippelius, \PRE{56,1998,R6275}.

\bibitem{ML98}
S. McNamara and S. Luding,  \PRE{58,1998,2247}.

\bibitem{LHMZ98}
S. Luding, M. Huthmann, S. McNamara, and A. Zippelius, \PRE{58,1998,3416}.

\bibitem{HHZ00}
O. Herbst, M. Huthmann, and A. Zippelius, \JL{Gran.\ Matt.,2,2000,211}.

\bibitem{AMZ01}
T. Aspelmeier, M. Huthmann, and A. Zippelius, in \emph{Granular Gases}, edited by T. P\"oschel and S. Luding (Springer, Berlin, 2001), pp.\ 31--58.

\bibitem{CLH02}
R. Cafiero, S. Luding, and H. J. Herrmann, \JL{Europhys.\ Lett.,60,2002,854}.

\bibitem{Z06}
A. Zippelius, \JL{Physica A,369,2006,143}.

\bibitem{L95}
S. Luding, \PRE{52,1995,4442}.

\bibitem{MHN02}
N. Mitarai, H. Hayakawa, and H. Nakanishi, \PRL{88,2002,174301}.


\bibitem{GNB05}
I. Goldhirsch, S. H. Noskowicz, and O. Bar-Lev, \PRL{95,2005,068002}.



\bibitem{BPKZ07}
N. V. Brilliantov, T. P\"oschel, W. T. Kranz, and A. Zippelius, \PRL{98,2007,128001}.

\bibitem{VT04}
P. Viot and J. Talbot, \PRE{69,2004,051106}.

\bibitem{PTV07}
J. Piasecki, J. Talbot, and P. Viot, \JL{Physica A,373,2007,313}.


\bibitem{CP08}
F. Cornu and J. Piasecki, \JL{Physica A,387,2008,4856}.


\bibitem{BDS97}
J. J. Brey, J. W. Dufty, and A. Santos, \JL{J.\ Stat.\ Phys.,87,1997,1051}.

\bibitem{VGS07}
F. Vega Reyes, V. Garz\'o, and A. Santos, \PRE{75,2007,061306}.






\end{thebibliography}
\end{document}